\documentclass[aps,prc,twocolumn,floatfix,amsmath]{revtex4}
\usepackage{dcolumn}
\usepackage{amssymb}
\usepackage{mathrsfs}
\usepackage{textcomp}
\usepackage{bm}
\usepackage{supertabular}
\usepackage{color,graphicx}
\usepackage[bookmarksopen]{hyperref}
\newcommand{\bea}{\begin{eqnarray}}
\newcommand{\eea}{\end{eqnarray}}
\newcommand{\nn}{\nonumber\\}

\begin{document}
 \title{Modification of heavy quark energy loss due to shear flow in hot QCD plasma}
\author{Sreemoyee Sarkar}
\email{sreemoyee.sarkar@saha.ac.in}

\affiliation{High Energy Nuclear and Particle Physics Division, Saha Institute of Nuclear Physics,
1/AF Bidhannagar, Kolkata-700 064, INDIA}

\medskip

\begin{abstract}
We present the derivation of heavy quark energy loss in a viscous QCD plasma using kinetic theory. Shear flow 
changes both
 boson and fermion distribution functions which eventually modify heavy quark energy loss. Due to presence 
of non-zero flow gradient in the medium all the bath particles here are out of equilibrium. In these types of 
plasmas we show that without plasma screening effects heavy quark energy loss suffers similar type of infrared divergence
 as one encounters 
in non-viscous plasma.
 The screening effects are incorporated consistently 
through Hard Thermal Loop resummation perturbation theory in the small-momentum-transfer region 
to obtain finite leading order result in $\eta/s$. 
We also quantify the importance of the result and demonstrate that shear flow has significant effect on the heavy quark energy loss. 
 
\end{abstract}

\maketitle
Recent years have witnessed significant progress in understanding the properties of hot and/or dense
 matter produced at relativistic heavy ion collider (RHIC) at BNL 
and the Large Hadron Collider (LHC)
 at CERN. Current research in this area has now generally accepted the fact that matter produced in these collider experiments
 behaves like nearly ideal fluid. Ideal hydrodynamics successfully estimates the 
  lower bound on shear viscosity ($\eta$) and entropy density ($s$) ratio ($\eta/s=1/4\pi$) of the
 fluid produced in the above mentioned experiments \cite{Kovtun05}.

Several other experimentally 
measured quantities such as
 elliptic flow ($v_2$) (as a function of $p_T$), particle type and impact parameter are also well 
described by ideal hydrodynamics. But there are some limitations to this 
success also. Ideal hydrodynamics fails to describe the trend of $v_2$ beyond $p_T\sim 2$ GeV. 
Above $p_T\sim 2$ GeV, $v_2$ does not rise as predicted by nonviscous
hydrodynamics. It also fails to describe certain relative trends observed in the 
baryon and meson elliptic flows \cite{Teaney03,Dusling10}. Current studies have been attributed 
in explaining these issues which reveal 
that invoking non-ideal viscous
 hydrodynamics falling trend of $v_2(p_T)$ in the higher $p_T$ region can naturally be explained. 
Incorporation of non-ideal effects into the theory eventually modifies
  stress-energy tensor ($T^{\mu\nu}$).
 Along with the ideal part it also receives viscous correction ($\delta T^{\mu\nu}$) \cite{Teaney03}. The modification of stress energy tensor due to non-zero flow gradient of the medium in turn
 modifies particle distribution function. The latter will now have a viscous part along with
   the non-viscous one ($f_0+\delta f$). The correction term $\delta f$ involves both the shear and the bulk viscosity coefficients and can be determined by solving Boltzmann
    equation. We however restrict ourselves only to the shear part \cite{Teaney03}.
    
Viscous corrected energy momentum
tensor and/or the distribution function modifies various experimental
     observables like particle spectra, Handbury Brown-Twiss radii,
or elliptic flow\cite{Soff01,Acdox02,Adler01}. Current studies on the modification of 
photon and dilepton spectra due to non-zero $\eta$ have explored the fact that in 
case of photon it leads to larger thermalization time. It has also been argued in \cite{Dusling10A,Bhatt10} that non-ideal effects 
increase net photon yield due to slowing down 
of hydrodynamic expansion. In case of dilepton the space-time integrated
transverse momentum spectra shows a hardening
where the magnitude of the correction increases with the
increasing invariant mass. In \cite{Dusling08} the authors argue that the
thermal description is reliable for an invariant mass $<2\tau_0T_0^2/(\eta/s)$, where, $\tau_0$ is the thermalization time
and $T_0$ is the initial temperature. Recently we have studied the effect of the shear flow on
 the fermionic damping rate \cite{AKDM13}, where it has been shown that like ideal 
Quantum Chromodynamic (QCD) plasma the magnetic sector remains logarithmic infrared divergent even after
 the incorporation of plasma screening effects through Hard Thermal Loop (HTL) mechanism. An attempt
has also been made to calculate the drag and diffusion
coefficients in viscous plasma numerically without considering plasma screening effects into the calculation 
\cite{Das12}. In the present work we present a consistent
 formalism of 
derivation of the leading order heavy quark energy loss in viscous QCD plasma with 
plasma screening effects 
into consideration. In the current work we restrict to the first order viscous correction 
upto $\cal{O}$ $(\eta/s)$ 
which allows us to present closed form analytical results. 

Since heavy quarks are good
   probe of QGP, several calculations have already been performed over the last decades to estimate 
the heavy quark energy loss ($-dE/dx$) in ideal plasma \cite{bjorken,thomaeloss1,thomaeloss2,svetitsky88,
moore05,dm05,sarkar10,sarkar11,Beraudo06,Peigne08a,Peigne08b}. Calculation of $-dE/dx$ 
in non-viscous QCD plasma has been plagued with infrared
    divergences. To deal with the problem, the usual way is to introduce Braaten and Yuaan's prescription where 
one separates the integration into two domains:
one involving the exchange of hard photons (or gluons),
i.e., the momentum transfer $q\sim T$ and the other involving
soft photons (or gluons) when $q\sim eT (gT)$ ($e, g\ll1$). In case
of the hard sector, one uses bare propagator and introduces
an arbitrary cutoff ($q^*$) parameter to regularize
the integration \cite{Yuan91}. For the latter, on the other hand, 
the hard thermal loop (HTL) corrected propagator is used. These
two domains, upon addition, yield results independent of the
intermediate scale \cite{Pisarski90,Braaten91}.
 
In all the previous works on the heavy quark energy loss the authors have 
 considered the fact that the bath particles are in equilibrium. In this paper we present a consistent formalism to calculate the heavy quark 
 energy loss with plasma screening effects where bath particles are affected by the longitudinal
 flow of the medium.  

 The problem of motion of a heavy quark in a QCD plasma looks familiar to that of a problem of test particle in 
plasma. The problem thus reduces to Brownian
  motion problem where quarks are executing random motion in plasma. To start with we appeal
 to the Boltzmann equation
\bea
\left(\frac{\partial}{\partial t}+{\bf v_p}.\nabla_{\bf r} +{\bf F}.\nabla_{{\bf p}}\right)f_p=-\mathcal{C}[f_p],
\label{boltz}
\eea
right hand side of the above equation represents the collision integral and $v_p=p/E_p$ is the velocity of the particle. In absence of external force and gradients of temperature, velocity or density on the injected parton, the above equation becomes,
\bea
\frac{\partial f_p}{\partial t}=-\mathcal{ C}[f_p].
\label{boltz_2}
\eea

In the present paper we consider a high energy heavy quark of mass $m_Q$ and momentum $p$ propagating through
 a QCD medium and scatters 
off the quarks and gluons of the bath. 
The heavy quark has energy $E_p$ and the mass of the light quarks
 in the bath $m_ q<<gT$. The injected heavy quark has 
a fluctuating part ($f_p=\delta f_p$) and all the bath particles are affected by the flow of the medium. The equilibrium part of the injected
 heavy quark vanishes since, $E_{p}>>T$. In the present work we are interested only in the $2\rightarrow 2$ ($P+K\rightarrow P'+K'$)
processes. The explicit form of the collision integral then becomes,
\bea
\mathcal{C}[f_p]&=&\frac{1}{2E_p}\int \frac{d^3k}{(2\pi)^3 2k}\frac{d^3p^{'}}{(2\pi)^3 2E_p'}\frac{d^3k^{'}}{(2\pi)^3 2k'}
 \delta f_pf_k(1\pm f_{k^{'}})
 \nn\
&\times&(2\pi)^4 \delta^4(P+K-P'-K')
\frac{1}{2}\sum_{spin}{\cal| M |}^2.
\label{collision_term}
\eea
Note the difference of the thermal phase space here with
that of the light quarks in \cite{dm05}. While writing the above equation for high energetic parton the possibility of 
back scattering has been excluded and the approximation $(1\pm f^0_{E_p'})\simeq 1$ has also been 
incorporated in the 
 thermal phase space since $E_{p'}>>T$.

The effect of flow of the medium is incorporated through the distribution function of the bath particles. We write 
viscous corrected distribution function as $f_i=f_i^0+\delta f_i^{\eta}$ ($\delta f_i^{\eta}<<f_i^0$), where, $i=k,k^{'}$. $\delta f_i^{\eta}$ is 
the first order correction to the thermal distribution function. 

Now, the expression for the energy loss can be obtained from Eqs.(\ref{boltz_2}) and 
(\ref{collision_term}) with the help of the relaxation
time approximation. With this approximation for the injected particle one writes,
\bea
\frac{\partial \delta f_p}{\partial t}=-\mathcal{ C}[ f_p]=-\delta f_p \Gamma(p).
\label{boltz_3}
\eea
$\Gamma(p)$ can be identified as the particle interaction rate.
The energy loss ($-dE/dx$) of heavy quark can be obtained by averaging over the interaction
rate times the energy transfer per scattering and dividing
by the velocity of the injected particle,
\bea
\frac{dE}{dx}=\frac{1}{v_p}\int d\Gamma \omega.
\label{e_loss_def}
\eea
With the help of the Eqs.(\ref{collision_term}), (\ref{boltz_3}) and (\ref{e_loss_def}) the heavy 
quark energy loss can be expressed as follows,
\bea
 &&-\frac{dE}{dx}\Bigg|(p)=\frac{1}{2E_pv_p}\int \frac{d^3k}{(2\pi)^3 2E_k}\frac{d^3p^{'}}{(2\pi)^3 2E_p'}\frac{d^3k^{'}}{(2\pi)^3 2E_k'}\nn\
  &\times&(f_k+\delta f_k^{\eta})(1\pm f_{k^{'}}\pm \delta f_{k'}^{\eta})\omega(2\pi)^4 \delta^4(P+K
 -P^{'}-K^{'})\nn\
&& \frac{1}{2}\sum_{spin}{\cal| M |}^2.
\label{eq:drag1}
\eea

 Now, the form of $\delta f_i^{\eta}$ mentioned above depends on the various ansatz \cite{Heiselberg93,
Heiselberg94,Arnold00,Dusling10},
\bea
\delta f_i^{\eta}=\chi(k)\frac{f_k^0(1\pm f_k^0)}{T}\hat k_i \hat k_j \nabla_i u_j.
\label{non_eq_dist_k}
\eea
In principle $\chi(k)$ can be determined from various microscopic theories as discussed in \cite{Dusling10}. 
In most of the hydrodynamic calculations it is assumed that $\delta f_k\propto k^2 f_k^0$ 
and the proportionality constant is independent of the particle type. 
This is known as quadratic ansatz \cite{Dusling10}.
For the present case we are interested in a boost invariant expansion without transverse flow. In this scenario
 one can incorporate the viscous correction to 
the distribution function in the following way \cite{Teaney03,Bhatt10,Das12},

 \bea
 \delta f^{\eta}_i(k)=f_i^0(1\pm f_i^0)\Phi_i(k),
 \label{vis_dist_func}
 \eea
 where,
\bea
\Phi_i(k)=\frac{1}{2T^3\tau}\frac{\eta}{s}\left(\frac{k^2}{3}-k_z^2\right).
\label{non_eqm_dist_func}
\eea
The viscous modification holds true only in the local rest frame 
 of the fluid and it contains the first order correction in the expansion of shear part
 of the stress tensor. $\tau$ is the thermalization time of the quark-gluon plasma (QGP) and the flow is 
along $z$ axis. From the
  above expression this is also evident that the non-equilibrium part of the distribution function 
  becomes operative only when there is a momentum anisotropy in the system. 
 
In a medium with non-zero flow gradient 
with the distribution functions mentioned in Eqs.(\ref{vis_dist_func}) and (\ref{non_eqm_dist_func})
the expression for energy loss can be written as,
\bea
&&-\frac{dE}{dx}\Bigg|(p)=\frac{A_q}{2E_pv_p}\int_{p',k,k'}\sum_{i=1,2}\alpha_i\omega(2\pi)^4\nn\ 
&&\delta^4(P+K
-P^{'}-K^{'})\frac{1}{2}\sum_{spin}{\cal| M |}^2
=\frac{-dE}{dx}\Bigg|^0+\frac{-dE}{dx}\Bigg|^{\eta},
\label{eloss}
\eea

 where, $\int_k$ is shorthand for $\int d^3k/(2\pi)^32E_k$ and $A_q=2n_f/3$ ($n_f$ is the number of flavor). In the above equation 
$\alpha_i$'s contain the information of the viscous modified phase-space factor. $\alpha_1$ 
 contains the equilibrium part of the distribution functions, this gives us the usual heavy quark energy loss
$\frac{-dE}{dx}|^0$ mentioned in
 \cite{bjorken,thomaeloss1,thomaeloss2,moore05,dm05,sarkar10,sarkar11,Beraudo06,Peigne08a,Peigne08b} 
where all the bath particles are in thermal equilibrium. $\frac{-dE}{dx}|^{\eta}$ is the
 viscous corrected energy loss. The equilibrium part of the phase space has the following form \cite{AKDM13},
\bea
\alpha_1=f_k^0(1\pm f_{k^{'}}^0),
\label{alpha_2}
\eea
and $\alpha_2$ involves terms due to the viscous modifications to the light quark distribution functions for the bath 
constituents \cite{AKDM13},
\bea
\alpha_2 &\simeq& \left[\Phi_kf_k^0(1\pm f_{k}^0)\pm \Phi_{k}f_{k^{'}}^0f_k^0\pm \Phi_{k^{'}}f_{k^{'}}^0f_k^0\right].
\label{alpha_3}
\eea
The above expression has been arrived at by neglecting terms ${\cal O}((\eta/s)^2)$ and ${\cal O}(f_i^3)$.

Inserting the above mentioned viscous corrected phase-space factor in Eq.(\ref{eloss}), one obtains,
\bea
&&-\frac{dE}{dx}\Bigg|^{\eta}(p)\simeq\frac{A_q}{2E_pv_p}\int_{k,p^{'},k^{'}} [\Phi_kf_k^0(1\pm f_{k}^0)\pm \Phi_{k}f_{k^{'}}^0f_k^0\nn\
&&\pm \Phi_{k^{'}}f_{k^{'}}^0f_k^0]
(2\pi)^4 
\delta^4(P+K-P^{'}-K^{'})\omega\frac{1}{2}\sum_{spin}|{\cal M}| ^2.
\label{damp_vis}
\eea

Since, in case of quark-quark (Q-q) scatterings small angle collisions give dominant contribution,  we write the phase-space factor with the
 following approximations,
\bea
&&f_{k^{'}}^0= f^0(k+\omega)\simeq f_k^0+\omega f_k^{0'}\nn\
&&\phi_{k^{'}}=\phi^0(k+\omega)\simeq \phi_k^0+\omega \phi_k^{0'}.
\eea
With the above expansion we approximate the phase-space factor upto ${\cal O}(f_i^2)$ and exclude 
higher order terms in $\omega$. Hence,
\bea
&&-\frac{dE}{dx}\Bigg|^{\eta}_{Qq}(p)\simeq\frac{A_q}{2E_pv_p}\int_{k,p^{'},k^{'}} [\Phi_kf_k^0(1- 3 f_{k}^0)
]\nn\
&&(2\pi)^4 
\delta^4(P+K-P^{'}-K^{'})\omega\frac{1}{2}\sum_{spin}|{\cal M}| ^2.
\label{damp_vis}
\eea
To proceed further we have to know the interaction. For $t$ channel Q-q scattering process the matrix element is 
given by  \cite{Peigne08a,Peigne08b},
\bea
\frac{1}{2}\sum_{spin}|{\cal M}|_{Qq} ^2&\propto&\frac{\tilde s^2}{t^2},
\label{mat_amp_bare_stu}
\eea
where, $\tilde s=s-m_Q^2$, $s$ and $t$ are the usual Mandalstam variables. 
%

Now, we consider the case of the hard gluon exchange where the medium effects on the propagator can be ignored.
 In this case one can see that $-\frac{dE}{dx}|^{\eta}_{Qq}$ is infrared divergent 
($-\frac{dE}{dx}|^{\eta}_{Qq}(p)\propto \int dq/q$) 
like non-viscous medium. The usual way to handle this divergences is
to incorporate the effects of plasma screening. The method
of calculating the effects of screening was developed by
Braaten and Yuan \cite{Yuan91}. As mentioned earlier this involves introduction of an arbitrary
momentum scale $q^*$ to distinguish the region of hard momentum
transfer from the soft region. The contribution from the hard momentum
region is calculated using tree-level scattering diagrams whereas HTL propagator is required for the soft
 momentum transfer.  The matrix
 amplitude with HTL resummed gluon propagator is necessary to evaluate  
$-\frac{dE}{dx}|^{\eta}_{Qq}$ in the latter domain. In the large wavelength limit $q<<T$
 this reduces to, 
\bea
\frac{1}{2}\sum_{spin}|\mathcal{ M}|^2_{Qq}& =&32g^4E_p^2k^2\Bigg[\frac{1}{\left(q^2+m_D^2\right)^2}\nn\
&+&
\frac{\left(v_p^2-\frac{\omega^2}{q^2}\right)q^2\mbox{cos}^2\phi}
{\left(1-\frac{\omega^2}{q^2}\right)\left(q^6+\frac{\pi^2\omega^2m_D^4}{16}\right)}\Bigg],
\label{mat_amp_cross_term}
\eea
where, $m_D\sim gT$ is the Debye mass. 
Evaluating both the hard and the soft sectors and restricting ourselves mainly to the leading logarithmic 
contribution one obtains,
\bea
-\frac{dE}{dx}\Bigg|^{\eta}_{Qq,l}(p)&\simeq&\left(\frac{\eta}{s}\right){\cal C}_1\left[f_1(v_p)-2f_1(v_p)\mbox{log}
\left|\frac{q_{max}}{m_D}\right|\right],\nn\
-\frac{dE}{dx}\Bigg|^{\eta}_{Qq,t}(p)&\simeq&\left(\frac{\eta}{s}\right)
{\cal C}_2\left[
\frac{4}{15}f_2(v_p)+f_2(v_p)\mbox{log}\left|\frac{2q_{max}}{\sqrt{\pi v_p}m_D}\right|\right],\nn\
\label{qq_t}
\eea
where $-\frac{dE}{dx}|_{Qq,l}^{\eta}$ and $-\frac{dE}{dx}|_{Qq,t}^{\eta}$ denote the longitudinal and the 
transverse contributions to the energy loss ($-\frac{dE}{dx}|_{Qq}^{\eta}
=-\frac{dE}{dx}|_{Qq,l}^{\eta}-\frac{dE}{dx}|_{Qq,t}^{\eta}$). 
The functions mentioned in the above equation have the following forms,  
\bea
{\cal C}_1&=&\frac{A_{q}g^4T}{(2\pi)^3\tau v_p^2}
\left(-\frac{7 \pi ^4}{60}+\frac{27 \text{Zeta}[3]}{2}\right), f_1(v_p)=\frac{v_p^5}{5}-\frac{v_p^3}{3},\nn 
{\cal C}_2&=&\frac{A_{q}g^4T}{(2\pi)^32\tau v_p^2}
\frac{4}{15}\left(-\frac{7 \pi ^4}{60}+\frac{27 \text{Zeta}[3]}{2}\right), f_2(v_p)=v_p^5,
\eea
where, $q_{max}$ in Eq.(\ref{qq_t}), can be approximated as 
$\sim \sqrt{E_p T}$ from the kinematics. 
 
To obtain $t$ channel contribution of quark-gluon (Q-g) 
scatterings from Eq.(\ref{qq_t}) the equation excluding $A_q$ has to be multiplied with the color factor 
$A_g=(N_c^2-1)/2=4$ ($N_c$ is the number of color). The total $t$ channel contribution to the 
$-dE/dx$ is then given by $-\frac{dE}{dx}|_{Q}^{\eta}=-(4+2n_f/3)\frac{dE}{dx}|_{Qq}^{\eta}$. In this regard
 it would be important to recall the expression for the ideal heavy quark energy loss (in $t$ channel) in \cite{thomaeloss2,moore05,Peigne08b}.
%

We now present the derivation of the contribution of Q-g scatterings to the heavy quark energy loss 
in $s$ and $u$ channels. One starts with the following expression, 
  \bea
   -\frac{dE}{dx}\Bigg|^{\eta}_{Qg}(p)&\simeq&\frac{A_fg^4}{v_p}\int \frac{d^3kd^3k'd^3p'}{(2\pi)^52E_p2E_{p'}2k2k'}
    \nn\
&\times&[\Phi_{k}f_k^0(1+f_{k}^0)
+\Phi_{k'}f_{k^{'}}^0f_{k}^0+\Phi_{k}f_{k^{'}}^0f_{k}^0]\nn\
&\times&\delta^4(P+K-P'-K')
 \left[\frac{-\tilde u}{\tilde s}+\frac{-\tilde s}{\tilde u}\right],
   \label{compt_vis_1}
   \eea
where, $\tilde u=u-m_Q^2$ and $A_f=16/9$. The $k'$ integration can be 
expressed as
\bea
&&\int_{k'} \frac{\Phi_{k}f_k^0(1+f_{k}^0)+
\Phi_{k'}f_{k^{'}}^0f_{k}^0+\Phi_{k}f_{k^{'}}^0f_{k}^0}{2k'}\,\nn\ 
&\times&(2\pi)^4
 \delta^{(4)}(P+K-P'-K')\nn\
&=&2\pi \Phi_k f_k^0(1+f_{k}^0)+
\Phi_{k+\omega}f_{k+\omega}^0f_{k}^0+
\Phi_{k}f_{k+\omega}^0f_{k}^0\nn\
&& \Theta({k+\omega})\,
\delta((K+Q)^2) .
\label{eq:k'-int}
\eea
The integration over $\phi_{p'}$ can be done with the help of the delta function as shown below,
\bea
\int_0^{2\pi} d\phi\, \delta((K+Q)^2) = \frac2{\sqrt{f}}\, \Theta(f) \, ,
\label{eq:phi-int}
\eea
where, $f=B^2-A^2$. $A$ and $B$ can be expressed in terms of the Mandelstam invariants 
\cite{Peigne08a,Peigne08b},
\bea
 A &=& s -m_q^2 +t-2kE_{p'}+2kp'\cos\theta_k\cos\theta_{p'} \, , \nn  \
 B &=& 2kp'\sin\theta_k\sin\theta_{p'} \, .
 \label{AB}
 \eea
  The variables can be changed from $p'$ and $\cos\theta_{p'}$ to $t$ and $\omega$ respectively by the following transformation, 
\bea
t &=& 2(m_Q^2 - E_pE_{p'} + p p' \cos\theta ), \nn\
\omega &=& E_p - E_{p'} \, .
\label{eq:change_variables}
\eea
With this, Eq.(\ref{compt_vis_1}) now becomes,
\bea
&&-\frac{dE}{dx}\Bigg|^{\eta}_{Qg}(p)\simeq\frac{A_fg^4}{16 \pi^2 v_ppE_p}\int_k \frac{1}{2k}\int_{-\infty}^0 dt
\int_{-\infty}^{\infty} \frac{d\omega \omega}{\sqrt{f(\omega)}}\,\nn\
&&\left( \Phi_k f_k^0(1+f_{k}^0)+
\Phi_{k+\omega}f_{k+\omega}^0f_{k}^0+
\Phi_{k}f_{k+\omega}^0f_{k}^0\right)g(s,t,\omega) \, ,
\label{eq:I}\nn\
\eea
 where, $g(s,t,\omega)$ depends on the Mandelstam variables and exchanged energy. 
Bounds on the integrals $\omega$ and $t$ arise from the condition $f=B^2-A^2\geq 0$. $f(\omega)$ 
can now be written as follows $f(\omega) = -a^2\omega^2 + b\,\omega +c$ \cite{Peigne08a,Peigne08b}.
The coefficients of the above equation are  \cite{Peigne08a,Peigne08b},
\bea
 a &=& \frac{s-m_Q^2}{p} \, , \nn  \
 b &=& -\frac{2t}{p^2}( E_p(s-m_Q^2) - k(s+m_Q^2) ) ,\nn \
 c &=& -\frac{t}{p^2} [ t( (E_p+k)^2-s ) + 4p^2k^2 -(s-m_Q^2-2E_pk)^2 ] .\nn\
 \label{abc}
 \eea
$f(\omega)$ is positive only in the domain $\omega_{min}<<\omega<<\omega_{max}$, where the discriminant $D=4a^2c+b^2$ is positive. 
Thus $\omega_{\rm min}^{\rm max}$ and $D$ can be evaluated along the line described in 
\cite{Peigne08a,Peigne08b}.
 The condition $D\geq 0$ leads to the $2\rightarrow 2$ scattering processes with one massless and one massive particle in the limit
  $t_{min}\leq t \leq 0$ with $ t_{\rm min} =-(s-m_Q^2)^2/s$ \cite{Peigne08a,Peigne08b}.
  Like Q-q scatterings here also we neglect terms which are more than ${\cal O}(f_i^2)$ 
and higher order in $\omega$. 

 Evaluation of the $\omega$ integral in Eq.(\ref{eq:I}) gives,
 \bea
I_\omega=\int_{\omega_{\rm min}}^{\omega_{\rm max}} d\omega\,\frac{\omega}{\sqrt{f(\omega)}}=
{\rm Re} \int_{-\infty}^\infty d\omega\, \frac{\omega}{\sqrt{f(\omega)}} \, =\frac{\pi b}{2 a^3}.\nn\
\eea
With the help of the above expression Eq.(\ref{eq:I}) reduces to,
\bea
&&-\frac{dE}{dx}\Bigg|^{\eta}_{Qg}(p)\simeq\left(\frac{\eta}{s}\right)\frac{A_fg^4}{4T^3\tau \pi v_p }\int_k 
\left(\frac{1}{3}-\cos^2\theta_{kz}\right)\nn\
&&\left(k^2f_k^0(1+3f_k^0)\right)
\left(1-\frac{(s+m_Q^2)k}{(s-m_Q^2)E_p}\right)
\int_{t_{min}}^0 dt\nn\
&&\frac{-t}{(s-m_Q^2)^2} \left[\frac{-\tilde u}{\tilde s}+\frac{-\tilde s}{\tilde u}\right]\, .
\label{eq:III}
\eea
As mentioned earlier we are interested in the energy loss of a high energetic parton 
$E_p>>m_Q^2/T$, which implies $s=m_Q^2+2PK\sim {\cal O}(E_pT)>>m_Q^2$. In this domain $\left(
(s+m_Q^2)k/(s-m_Q^2)E_p\right)\rightarrow 0$. 
Finally the expression for the Q-g scatterings in the $s$ and the $u$ channel reduces to,
\bea
&&-\frac{dE}{dx}\Bigg|^{\eta}_{Qg}(p)\simeq\left(\frac{\eta}{s}\right){\cal C}_3
\Bigg[\frac{-11T^4}{18}\left(18 \zeta (3)-\frac{2 \pi ^4}{15}\right)\nn\
&-&\frac{T^3m_Q^2}{3E_p}\left
(\left(\pi ^2-4 \zeta (3)\right)\ln\left|\frac{4E_pT}{m_Q^2}\right|-.162225\right)\Bigg],
\label{qg_su}
\eea
 ${\cal C}_3=A_fg^4/(32T^3\tau \pi^3 v_p)$. The leading logarithmic term of Q-g scatterings 
(both in s and u channels) in non-viscous medium is given in \cite{Peigne08b}. 
 The final expression of heavy quark energy energy loss can be obtained by adding Eqs.(\ref{qq_t}),
 and (\ref{qg_su}) along with the ideal contributions,
\bea
&&-\frac{dE}{dx}\Bigg|(p)=\frac{-dE}{dx}\Bigg|^{0}_{Qq}+
\frac{-dE}{dx}\Bigg|^{0}_{Qg}+\frac{-dE}{dx}\Bigg|^{\eta}_{Qq}+\frac{-dE}{dx}\Bigg|^{\eta}_{Qg}\nn\
&=&\frac{-dE}{dx}\Bigg|^{0}_{Qq}+
\frac{-dE}{dx}\Bigg|^{0}_{Qg}+\left(\frac{\eta}{s}\right)\left(4+\frac{2n_f}{3}\right)\Bigg(-\frac{7 \pi ^4}{60}\nn\
&+&\frac{27 \text{Zeta}[3]}{2}\Bigg)
\Bigg[\frac{g^4T}{(2\pi)^3\tau v_p^2}
\left\{f_1(v_p)-2f_1(v_p)\mbox{log}
\left|\frac{q_{max}}{m_D}\right|\right\}\nn\
&+&\frac{g^4T}{60\pi^3\tau v_p^2}
\left\{
\frac{4}{15}f_2(v_p)+f_2(v_p)\mbox{log}\left|\frac{2q_{max}}{\sqrt{\pi v_p}m_D}\right|\right\}\nn\
&+&\left(\frac{\eta}{s}\right)\frac{A_fg^4}{32T^3\tau \pi^3 v_p}
\Bigg\{\frac{-11T^4}{18}\left(18 \zeta (3)-\frac{2 \pi ^4}{15}\right)\nn\
&-&\frac{T^3m_Q^2}{3E_p}\left
(\left(\pi ^2-4 \zeta (3)\right)\ln\left|\frac{4E_pT}{m_Q^2}\right|-.162225\right)\Bigg\}\Bigg].
 \eea
 The above expression 
for energy loss in a plasma where all the bath particles are affected by the shear flow of the medium reveals
 the fact that similar to the heavy quark energy loss in a non-viscous medium this also suffers from infrared divergence for bare propagation. Using
 HTL propagator we obtain closed form leading order analytic result first order in viscous correction ${\cal O}(\eta/s)$.
\begin{figure}[t]
\begin{center}
\includegraphics[scale=0.33]{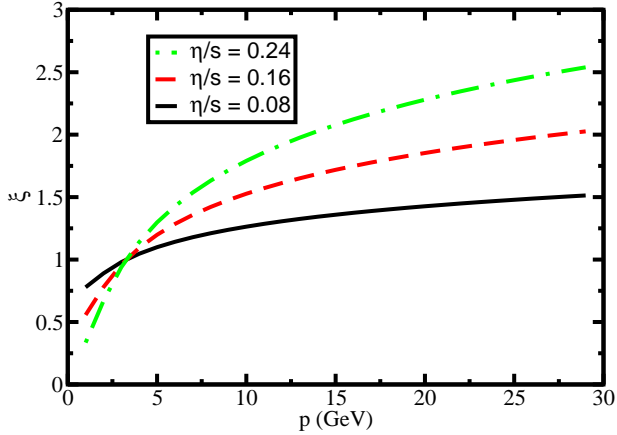}
\end{center}
\vspace{-0.4cm}
\caption{(Color online)  The variation of $\xi$ of charm quark with momentum at 
$T= 0.225$ GeV, $\tau=0.3$ fm, $m_D=0.5$ GeV and $\alpha=0.3$.}   
\label{eloss_p}
\end{figure}
\begin{figure}[!htb]
 \begin{center}
 \includegraphics[scale=0.33]{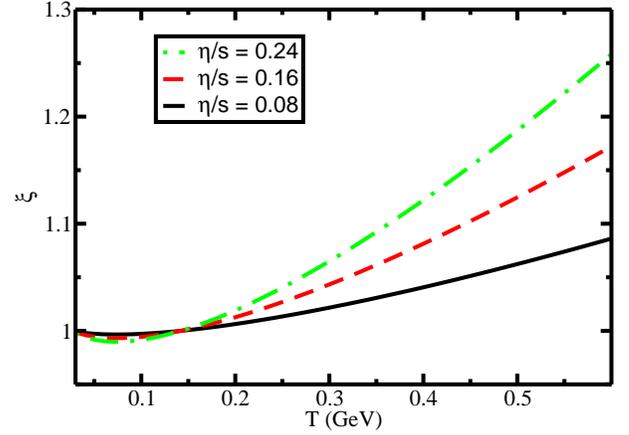}
 \end{center}
 \vspace{-0.4cm}
 \caption{(Color online) The variation of $\xi$ of charm quark with temperature at $p=5$ GeV, $\tau=0.3$ fm, $m_D=0.5$ GeV and $\alpha=0.3$.}   
 \label{eloss_T}
 \end{figure}
 
To quantify the nature of heavy quark energy loss we here plot fractional energy loss (
$\xi=\frac{dE}{dx}/\frac{dE}{dx}|^0$) with momentum ($p$) in Fig.(\ref{eloss_p}) and with temperature ($T$) in 
Fig.(\ref{eloss_T}) for charm quark.
In Fig.(\ref{eloss_p}) we observe the momentum variation of $\xi$ for charm quark. From the plot it is clear that 
at $T=0.225$ GeV and at low momentum $p<5$ GeV increase in $\eta/s$ decreases the relative energy loss, whereas in the high momentum
 region $p>5$ GeV opposite nature can be seen. Fig.(\ref{eloss_T}) depicts the variation 
 of $\xi$ with temperature at $p=5 $ GeV. In Fig.(\ref{eloss_T}) note that at lower temperature
 regime $\xi$ decreases and after $T=0.15$ GeV it starts to increase. Hence, 
there is a crossover of trend of $\xi$ with $\eta/s$ at $T\sim 0.15$ GeV. The nature of the two plots can be understood from the flow modified phase space factor.
 Since, the anisotropic phase-space part is negative in the low momentum/temperature region this gives negative contribution to the total energy loss whereas in the higher 
 momentum/temperature
 region this contributes positively. 

To summarize, in the present work we have calculated heavy quark energy loss in a medium where all the bath 
particles are affected by the shear flow of the medium. It has been shown in the text 
that shear viscosity enters into the 
calculation through the viscous corrected phase-space factor and this modifies the result of the energy loss. 
In the present work only elastic Q-q, 
Q-g scatterings have been considered and the leading order correction terms in $\eta/s$ to the energy loss have been 
estimated. We observe that in case of viscous medium also bare gluon propagation gives rise to 
logarithmic infrared
 divergent $\int dq/q$ result similar to the case of non-viscous medium. HTL resummed propagator is used to circumvent the problem. While performing
 the calculation only small angle contributions are considered since these provide us 
the dominant contribution to the energy loss.
 Along with the above mentioned points it has also been considered that only longitudinal shear flow is present in
 the current derivation. 
The approximations mentioned above allow us to present
 leading logarithmic, first order correction term in $\eta/s$ of the heavy quark energy loss.
 Moreover, one of the interesting findings
 of the present work is, the plasma effects in a viscous medium increases heavy quark energy loss in the high momentum
 region ($p>5$ GeV). In the low temperature region ($T<0.15$ GeV) the energy loss shows opposite trend in comparison to
 the high temperature one. Findings of the present work will have significant consequences in studying different 
observables
 like nuclear modification factor, particle spectra in recent heavy ion collision experiments.

 I am indebted to Late Prof. A. K. Dutt-Mazumder who introduced me to this topic and his fruitful discussions motivated me to initiate the present work. 
I would also like to acknowledge Prof. J. P. Blaizot, Prof. P. K. Roy and Dr. S. K. Das for their valuable discussions and comments.


\begin{thebibliography}{50}
\bibitem{Kovtun05} P. K. Kovtun, D. T. Son and A. O. Starinets 
Phys. Rev. Lett. {\bf 94}, 111601 (2005).
\bibitem{Teaney03} D. Teaney, Phys. Rev. C {\bf 68}, 034913 (2003).
\bibitem{Dusling10} K. Dusling, G. D. Moore and D. Teaney, Phys. Rev. C {\bf 81}, 034907 (2010).

\bibitem{Soff01} S. Soff, S. A. Bass and Adrian Dumitru, Phys. Rev. Lett. {\bf 86}, 3981 (2001).
\bibitem{Adler01} C. Adler {\em et al.}, STAR Collaboration, Phys. Rev. Lett. {\bf 87}, 082301 (2001).
\bibitem{Acdox02} K. Acdox {\em et al.}, PHENIX Collaboration, Phys. Rev. Lett. {\bf 88}, 192302 (2002).
\bibitem{Dusling10A} K. Dusling Nucl. Phys. A. {\bf 839}, 70 (2010).

\bibitem{Bhatt10} J. R. Bhatt, H. Mishra and V. Sreekanth, JHEP {\bf 11}, 106 (2010).
\bibitem{Dusling08} K. Dusling Nucl. Phys. A. {\bf 809}, 245 (2008).
\bibitem{AKDM13} S. Sarkar and A. K. Dutt-Mazumder, Phys. Rev. D {\bf 88}, 054006 (2013).
\bibitem{Das12} S. K. Das, V. Chandra and Jan-e Alam, J. Phys. G {\bf 41}, 015102 (2013).

\bibitem{bjorken} D. Bjorken, Fermilab Report No. PUB-82/59-THY,
1982 (unpublished).
\bibitem{thomaeloss1} E. Braaten and M.H. Thoma, Phys. Rev. D {\bf 44}, 1298(1991).
 \bibitem{thomaeloss2} E. Braaten and M.H. Thoma, Phys. Rev. D {\bf 44}, R2625(1991).
 \bibitem{svetitsky88} B. Svetitsky, Phys. Rev. D {\bf 37}, 2484(1988).
 \bibitem{moore05} G. D. Moore, D. Teaney Phys. Rev. C {\bf 71}, 064904(2005).
 \bibitem{dm05} A.K.Dutt-Mazumder, Jan-e Alam, P. Roy and B. Sinha, Phys. Rev. D {\bf 71}, 094016(2005).
 \bibitem{sarkar10} S. Sarkar and A. K. Dutt-Mazumder, Phys. Rev. {\bf D82},  056003  (2010).
 \bibitem{sarkar11} S. Sarkar and A. K. Dutt-Mazumder, Phys. Rev. {\bf D84},  096009  (2011).
\bibitem{Beraudo06} A. Beraudo, A. De Pace, W.M. Alberico, A. Molinari, Nucl. Phys. A {\bf 831}, 59(2009).
\bibitem{Peigne08a} S. Peigne and A. Peshier, Phys. Rev. D {\bf 77}, 014015 (2008).
\bibitem{Peigne08b} S. Peigne and A. Peshier, Phys. Rev. D {\bf 77}, 114017 (2008).
\bibitem{Yuan91} E. Braaten and T. C. Yuan, Phys. Rev. Lett. {\bf 66}, 2183(1991).
\bibitem{Pisarski90} E. Braaten and R. D. Pisarski, Phys. Rev. Lett. {\bf 64}, 1338(1990).
\bibitem{Braaten91} E. Braaten and R. D. Pisarski, Nucl. Phys. B {\bf 337}, 569(1970).
 \bibitem{Heiselberg93}  H. Heiselberg and C. J. Pethick, Phys. Rev. D. {\bf 48}, 2916(1993).
 \bibitem{Heiselberg94}  H. Heiselberg, Phys. Rev. D. {\bf 49}, 4739(1994).
\bibitem{Arnold00} P. Arnold, G. D. Moore and L. G. Yaffe JHEP {\bf 0011}, 001(2000).
\bibitem{Groot} S. de Groot, W. van Leevuen and Ch. van Veert, {\em Relativistic Kinetic Theory} (North-Holland, Amsterdem, 1980).
\bibitem{Arnold03} P. Arnold, G. D. Moore and L. G. Yaffe JHEP {\bf 0305}, 051(2003).

\end{thebibliography}
\end{document}